\documentclass[aps,twocolumn,floatfix,nofootinbib,superscriptaddress]{revtex4}
\usepackage{subfigure}
\usepackage{eucal}
\usepackage{psfrag,subfigure}
\usepackage{graphicx,psfrag,subfigure}
\def\prb{Phys. Rev. B}
\def\prl{Phys. Rev. Lett.}

\def\be{\begin{equation}}
\def\ee{\end{equation}}
\def\ba{\begin{eqnarray}}
\def\ea{\end{eqnarray}}

\def\LCO{La$_2$CuO$_4$}

\def\LBCO{La$_{2-x}$Ba$_x$CuO$_4$}
\def\LBCOc{La$_{15/8}$Ba$_{1/8}$CuO$_4$}

\def\C60{A$_x$C$_{60}$}
\def\LNSCO{La$_{1.6-x}$Nd$_{0.4}$Sr$_x$CuO$_{4}$}

\def\HgCu3{HgCa$_2$Cu$_3$O$_{8+y}$}
\def\HgCu4{HgBa$_2$Ca$_3$Cu$_4$O$_{10+y}$}
\def\TlCu{Tl$_2$Ba$_2$CuO$_{6+\delta}$}
\def\TlCu3{Tl$_2$Ba$_2$Ca$_2$Cu$_3$O$_{10+y}$}
\def\TlCu4{Tl$_2$Ba$_2$Ca$_3$Cu$_4$O$_{12+y}$}

\def\BiCu3{Bi$_2$Sr$_2$Ca$_{2}$Cu$_3$O$_y$}

\def\C60{A$_x$C$_{60}$}

\begin{document}

\title{
Dynamical layer decoupling in a stripe-ordered, high 
$T_c$ superconductor 
 }
\author{
E. Berg}
\affiliation{Department of Physics, Stanford University, Stanford, California 94305-4060}
\author{
E. Fradkin}
\affiliation{Department of Physics, University of Illinois at Urbana-Champaign,  
Urbana, Illinois 61801-3080}
\author{
E.-A. Kim}
\affiliation{Department of Physics, Stanford University, Stanford, California 94305-4060}
\author{
S. A. Kivelson}
\affiliation{Department of Physics, Stanford University, Stanford, California 94305-4060}
\author{
V. Oganesyan}
\affiliation{Department of Physics, Yale University, New Haven, Connecticut 06520-8120}
\author{
J. M. Tranquada}
\affiliation{
Brookhaven National Laboratory, Upton, New York 11973-5000}
\author{
S. C. Zhang}
\affiliation{Department of Physics, Stanford University, Stanford, California 94305-4060}
\date{\today}
\begin{abstract}
In the stripe-ordered state of a strongly-correlated two-dimensional electronic system, under a set of special circumstances, the superconducting condensate, like the magnetic order, can occur at a non-zero wave-vector corresponding to a spatial period double that of the charge order.  In this case, the Josephson coupling between near neighbor planes, especially in a crystal with the 
special
structure of {\LBCO}, vanishes identically.  We propose that this is the underlying cause of the dynamical decoupling of  the layers
recently observed in transport measurements at $x=1/8$.
\end{abstract} 
\pacs{}
\maketitle

High-temperature superconductivity (HTSC) was first discovered \cite{bandm} in 
{\LBCO}.
%Moreover, it 
%It exhibits a
A
sharp anomaly \cite{modenbaugh} in $T_c(x)$ occurs at $x=1/8$ 
which is now known to
be indicative \cite{tranquada,abbamonte} of
the existence of stripe order and of its strong interplay with HTSC.
%Most recently, a 
%SAK truly 
Recently, a remarkable %and unexpected 
dynamical layer decoupling has been observed \cite{newtranq} 
associated with the superconducting (SC) fluctuations 
%primarily 
below the spin-stripe ordering transition temperature, $T_{\rm spin}=42$K.  

While $T_c(x)$, as determined by the onset of a bulk
Meissner effect,  reaches values up to $T_c(x=0.1)=33$~K
%SAK32 
%33 K$ 
for $x$ somewhat smaller and larger than 
$x=1/8$, 
%as determined by the onset of a bulk
%Meissner effect, 
$T_c(x)$ drops to 
 the range 2--4~K for $x=1/8$.  However, in
other respects, superconductivity appears to be optimized for $x=1/8$.
%SAK;  the 
The d-wave gap determined by ARPES has recently been shown \cite{valla} to
be largest for $x=1/8$.  Moreover, strong SC fluctuations
produce an order of magnitude drop \cite{newtranq} in the in-plane
resistivity, $\rho_{ab}$, at  $T\approx T_{\rm spin}$, which is %a
considerably higher %temperature
 than  the highest bulk SC.
%$T_c(x=0.1)=33$~K.  

%Still 
%more dramatically, the f
The fluctuation conductivity reveals heretofore
unprecedented characteristics (%shown 
as described schematically in Fig.~1):  1) 
$\rho_{ab}$ drops rapidly with decreasing temperature from $T_{\rm spin}$
down to $T_{KT}\approx 16$K, at which point it becomes unmeasurably
small.  
In the range $T_{\rm spin} > T > T_{KT}$,  the
temperature dependence of $\rho_{ab}$ is qualitatively of the Kosterlitz-Thouless form,
as if the
SC fluctuations were strictly confined to a single
copper-oxide plane.  2)  By contrast, the resistivity perpendicular to
the copper-oxide planes, $\rho_{c}$, increases with decreasing
temperatures from $T^\star \gtrsim 300$~K, down to $T^{\star\star}\approx
35$~K.  For $T < T^{\star\star}$, $\rho_c$ decreases with decreasing
temperature, but it only becomes vanishingly small below $T_{3D}\approx
10$~K.
Within experimental error, for
$T_{KT} > T > T_{3D}$, the 
%SAKratio of resistivities
resistivity ratio, $\rho_{c}/\rho_{ab}$, is infinite!  
3)  The full set of usual characteristics of the SC state,
the Meissner effect and perfect conductivity, $\rho_{ab}=\rho_c=0$, is
only observed below $T_c=4K$.  Thus, for $T_{3D} > T > T_c$, a peculiar
form of fragile
3D superconductivity exists.  

\begin{figure}[ht!!!]
\includegraphics[width=0.44\textwidth]{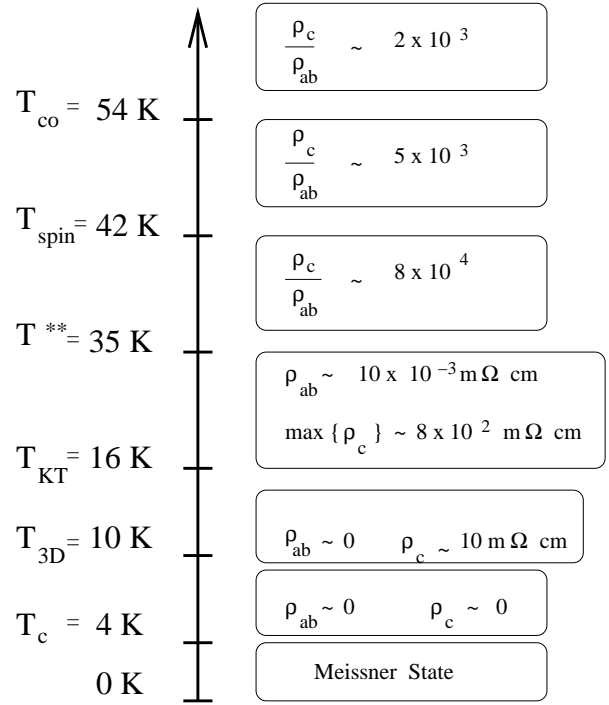}
\caption{  Summary of the thermal phase transitions and transport regimes in $x=1/8$ doped {\LBCO}.}
\end{figure}

The above listed results are new, so 
an
extrinsic explanation of 
some  aspects of the data is possible.  
Here we assume that the measured properties
do
reflect the bulk behavior of {\LBCO}.  We show that there is
a straightforward way in which stripe order can lead to an enormous
dynamical suppression of interplane Josephson coupling, particularly in
the 
charge ordered low-temperature tetragonal
(LTT) 
phase of
{\LBCOc}, {\it i.e.\/} 
$T \leq T_{\rm co}=54$ K.  

The LTT structure has two planes per unit cell.  In alternating planes,
the charge stripes run along the x or y axes, as shown in Fig. 3. 
Moreover, the parallel stripes in second neighbor planes are thought to
be shifted over by half a period (so as to minimize the
Coulomb interactions \cite{zimmerman}) resulting in a further doubling of the number of
planes per unit cell, as seen in X-ray scattering studies.  Below
$T_{\rm spin}$, the spins lying between each charge stripe have antiferromagnetic (AFM) order along the stripe direction,
which suffers a $\pi$
phase shift across each charge stripe, resulting in a doubling of the
unit cell within the plane, 
see Fig. 2c.  
Hence,  the
Bragg scattering from the charge order in a given plane occurs at
$(2\pi/a)\langle \pm 1/4,0\rangle$ while the spin-ordering occurs at 
$(2\pi/a)\langle 1/2\pm 1/8,1/2\rangle$.  

 SC order
 %, as it
%develops,
should 
occur
most strongly within the charge stripes.   Since
%SC order 
it is strongly associated with zero 
center-of-mass
momentum
pairing, one usually expects, and typically finds
in models, 
that the
SC order on neighboring stripes has the same phase. 
However, as we will discuss, under special circumstances, the
SC order, like the AFM order, may suffer a
$\pi$ phase shift between neighboring stripes if the effective Josephson
coupling between stripes is negative.  Within a plane, so long as the
stripe order is defect free, the fact that the  SC order
occurs with $k=(2\pi/a)\langle\pm 1/8,0\rangle $ has only limited
observable consequences.  
However
%, as can readily be seen, 
anti-phase SC order within a
plane results in an exact cancellation of the effective Josephson coupling
between first, second and third neighbor planes.  This observation can
explain an enormous 
reduction of the
interplane
SC correlations in a stripe-ordered phase.  

Before proceeding, we 
%want to 
remark that there is a preexisting observation, concerning the spin order, which supports the idea that interplane decoupling is a bulk feature of a stripe-ordered phase.  Specifically, although the in-plane spin correlation length measured in neutron-scattering studies in particularly well prepared crystals of {\LBCO} is  $\xi_{\rm spin}\geq 40 a$ \cite{fujita}, there are essentially no detectable 
magnetic correlations between neighboring planes. In typical circumstances, 3D ordering would be expected to onset when $(\xi_{\rm spin}/a)^2 J_1 \sim T$, where $J_1$ is the strength of the interplane exchange coupling.  However, the same geometric frustration of the interplane couplings that we have discussed in the context of the SC order pertains to the magnetic case, as well.  Thus, we propose that the same dynamical decoupling of the planes is the origin of both the extreme 2D character of the AFM and SC ordering.

We begin with a caricature of a stripe ordered state, consisting of
alternating Hubbard or $t$--$J$ ladders which are weakly coupled to each
other (Fig.~2).  Such a caricature, which has been adopted 
in previous studies of superconductivity in stripe ordered
systems \cite{ekz,afk,getal}, certainly overstates the extent to which
stripe order produces quasi-1D electronic structure.  However, %we believe that 
we can learn something about the possible electronic phases
and their microscopic origins, in the sense of adiabatic continuity, by
analyzing the problem in  this extreme limit.  As shown in the figure,
distinct patterns of period 4 stripes can be classified by their pattern
of point group symmetry breaking as being 
%SAK
``bond centered'' or ``site centered.'' 
Numerical studies of $t$--$J$ ladders \cite{8leg} suggest that the difference
in energy between bond and site centered stripes is small, so the balance
could easily be tipped one way or another by material specific details,
such as the specifics of the electron-lattice coupling.
%%%%%%%%%%%%%%%%%%%%%%%%%%%%%
\begin{figure}[t]
\subfigure[~Bond centered]{
\includegraphics[width=.2\textwidth]{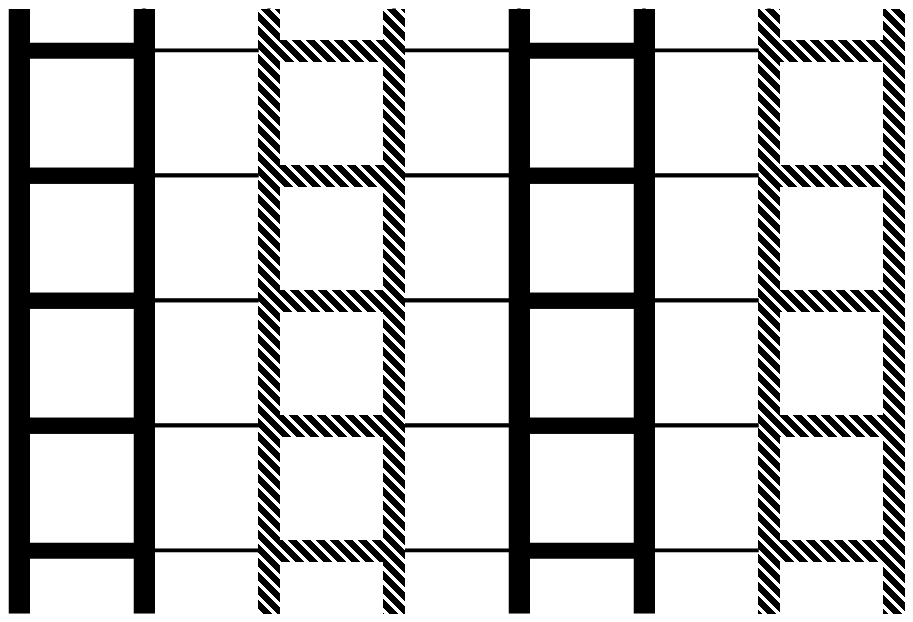}
}
\subfigure[~Site centered]{
\includegraphics[width=.2\textwidth]{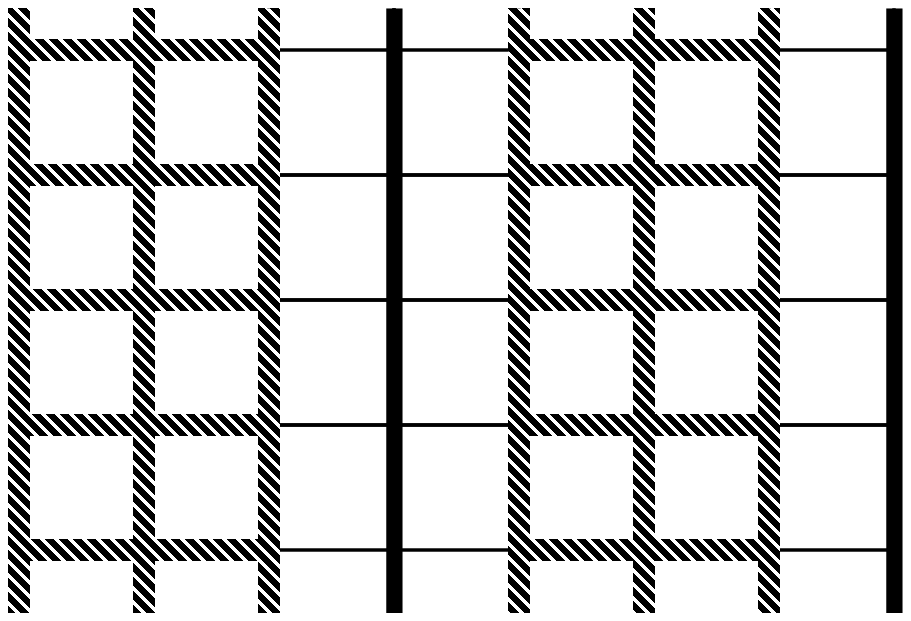}
}\\
\subfigure[~Magnetic striped]{
\includegraphics[width=.25\textwidth]{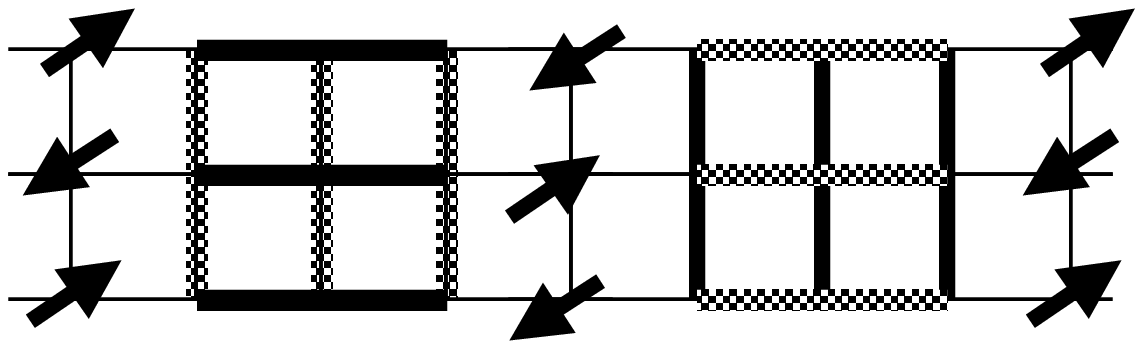}
}
\caption{a) Pattern of a period 4 bond centered and b)  site centered
stripe, with nearly undoped (solid lines) and  more heavily doped (hatched
lines) regions.  c) Sketch of the pair-field
(lines) and spin (arrows) order in a period 4 site centered stripe in
which
both the SC and AFM order have
period 8 due to an assumed  $\pi$ phase shift across the intervening
regions.  Solid (checked) lines denote a positive (negative) 
pair-field.}
\end{figure}
%%%%%%%%%%%%%%%%%%%%%%%%%%%%%%%%
%%%%%%%%%%%%%%%%%%%%%%%%%%%%%%%%
\begin{figure}[b]
   \includegraphics[width=.25\textwidth]{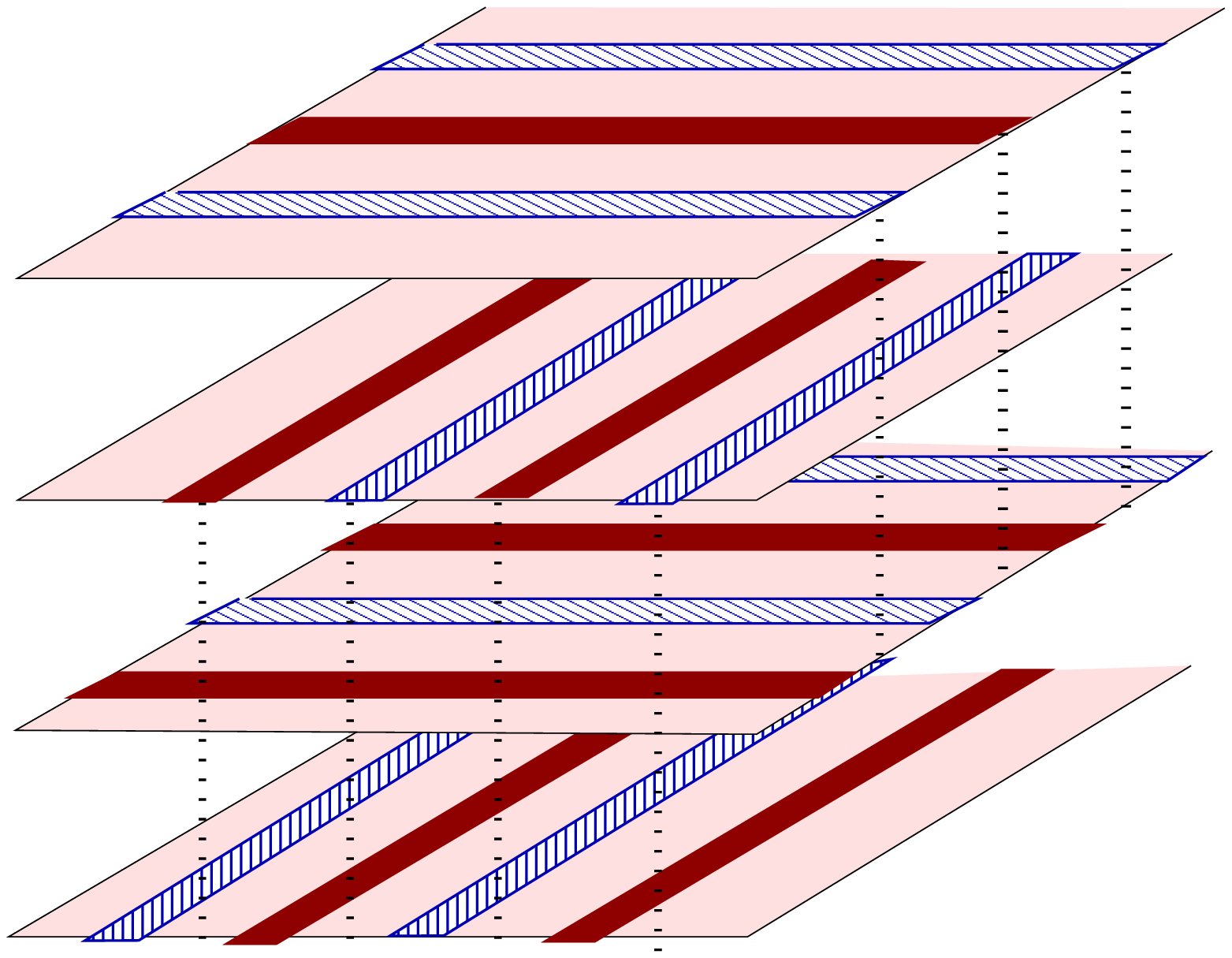} 
   \caption{Stacking of stripe planes.}
\end{figure}
%%%%%%%%%%%%%%%%%%%%%%%%%%%%%%%%%%

The simplest caricature of bond centered stripes is an array of weakly
coupled two-leg ladders with alternately larger and smaller doping, as
illustrated in Fig.~2a.   This problem was studied in Ref.
\onlinecite{afk}.   Because a strongly interacting electron fluid on a
two-leg ladder  readily develops a spin-gap,\cite{white2leg} {\it i.e.}
forms a LE liquid, this structure can exhibit strong
SC tendencies to high temperatures.  Weak electron hopping
between neighboring ladders produces Josephson coupling which can lead to
a ``d-wave like'' SC state.\cite{dwave} 
However, the spin-gap precludes any
form of magnetic ordering, even when the ladders are weakly coupled, 
and there is nothing about the SC order that would
prevent phase locking between neighboring planes in a 3D material.  For
both these reasons, this is not an attractive model for the stripe
ordered state in {\LBCOc}. (There is, however, evidence from
STM studies on the surface of BSCCO \cite{davis} of self-organized
structures suggestive of two-leg ladders.)  
 
By contrast, a site-centered stripe is naturally related to an
alternating array of weakly coupled three and one leg ladders, as shown
in Fig.~2(b).  Because the zero-point kinetic energy of the doped holes is
generally large compared to the exchange energy, it is the three-leg
ladder that we take to be the more heavily doped.  The three leg ladder
is known \cite{ekz,3leg} to develop a spin-gapped LE liquid
above a rather small \cite{3leg} critical doping, $x_c$ (which depends on
the 
%strength of the 
interactions).  An undoped or lightly doped one-leg
ladder, by contrast, is better thought of as an incipient spin density
wave (SDW), and has no spin-gap.  Where the one-leg ladder is lightly
doped it forms a Luttinger liquid with a divergent SDW susceptibility at
$2k_F$.  The phases of a system of alternating, weakly coupled
LE and Luttinger liquids 
%SAKwas 
were analyzed in 
\cite{getal}.  However, the magnetic order in {\LBCOc} produces a
Bragg peak at wave-vector $(\frac{2\pi}{a})\langle \frac{1}{2}\pm \frac{1}{8}, \frac{1}{2} \rangle$ in a
coordinate system in which $y$ is along the stripe direction. 
Therefore, it is necessary to consider the case in which, in the absence
of inter-ladder coupling, the one-leg ladder is initially undoped, and the
three leg ladder has $x= \frac{1}{6}> x_c$.

Our model of the electronic structure of a single charge-stripe-ordered Cu-O plane is thus an alternating array of LE liquids, with a spin-gap but no charge gap, and spin-chains, with a charge gap but no spin gap.  None of the obvious couplings between nearest-neighbor subsystems is relevant in the renormalization group sense, because of the distinct character of their ordering tendencies.  However,  certain induced second neighbor couplings, between identical systems, are strongly relevant, and, at 
$T=0$, lead to a broken symmetry ground-state.  

The induced exchange coupling between nearest-neighbor spin-chains leads to
a 2D magnetically ordered state.  The issue of the sign of this coupling has
been addressed previously \cite{zachar,pryadko,liu01} and found to be
non-universal, as it depends on the doping level in the
intervening three-leg ladder.  For $x=0$,  the preferred 
AFM order is in-phase on neighboring spin-chains,
consistent with a magnetic ordering vector of $(2\pi/a)\langle
1/2,1/2\rangle$.  For large enough $x$ (%which 
probably, %means any 
$x>x_c$),
the ordering on neighboring spin-chains is $\pi$ phase shifted, resulting
in a doubling of the unit-cell size in the direction perpendicular to the
stripes, and a magnetic ordering vector $(2\pi/a)\langle 1/2\pm 1/8, 1/2
\rangle$.  
This ordering
tendency 
has also been %documented directly 
found in studies of
wide $t$--$J$ ladders \cite{8leg}.  

A question that 
has not  been addressed systematically until now is the sign of the effective Josephson coupling between neighboring LE liquids.  In the case of 2-leg ladders, it was found \cite{afk,8leg} that the effective Josephson coupling is positive, favoring a SC state with a spatially uniform phase. % However, it 
It is possible, in highly correlated systems, especially when tunneling through a magnetic impurity \cite{spivak}, to encounter situations in which the effective Josephson coupling is negative, therefore producing a $\pi$-junction.  %Some time ago, 
Zhang \cite{zhang} has
observed that, regardless the microscopic origin of the  anti-phase character of the magnetic ordering in the striped state,
if there is an approximate SO(5) symmetry relating the antiferromagnetism to the superconductivity, one should expect an 
anti-phase ordering of the superconductivity in a striped state.  
The example of tunneling through decoupled magnetic impurities \cite{spivak} 
is a proof in principle that such behavior {\it can} occur. 
However, 
%SAKdramatic 
 interplane decoupling associated with the onset of superconductivity is not 
 %SAKclearly 
 seen in experiments in other cuprates, and states with periodic $\pi$ phase shifts of the SC order parameter have not yet surfaced  in numerical studies of microscopic models \cite{8leg};  this suggests anti-phase striped SC order is rare.

The %principal 
new proposal in the present paper is that, for the reasons outlined above, the SC striped phase of {\LBCOc} has anti-phase SC 
%as well as 
and anti-phase AFM order, whose consequences we now outline. 
%Let us now outline some of the consequences of this proposition. 
We can express the most important possible interplane Josephson-like coupling terms compactly as
\begin{eqnarray}
H_{\rm inter} = && \sum_j \int d\vec r\sum_{n,m}{\cal J}_{n,m} \left [\left( \Delta_j^\star\Delta_{j+m}\right)^n + {\rm h.c.}\right] 
\end{eqnarray}
where $\Delta_j$
 is the $j$-th plane SC order parameter. %on plane $j$.  
The term proportional to the  usual (lowest order) Josephson coupling, ${\cal J}_{1,1}$, and indeed,  ${\cal J}_{1,2}$ and  ${\cal J}_{1,3}$ all vanish by symmetry.  The most strongly relevant residual interaction is the Josephson coupling between fourth-neighbor planes, ${\cal J}_{1,4}$.  
%More weakly relevant, but with a probably larger bare value, is 
Double-pair
 tunnelling between nearest-neighbor planes, ${\cal J}_{2,1}$, is more weakly relevant, but it probably has a larger bare value since it 
involves half as many powers of the single-particle interplane matrix elements than ${\cal J}_{1,4}$.  
${\cal J}_{1,4}$ and ${\cal J}_{2,1}$ have scaling dimensions $1/4$ and $1$ at the (KT) critical point 
of decoupled plains, so both are relevant.  
Thus, they become important 
when the in-plane SC correlation length $\xi
\sim \xi_{1,4} \sim [{\cal J}_o/{\cal J}_{1,4}]^{1/4}$ and $\xi_{2,1} \sim [{\cal J}_o/{\cal J}_{2,1}]$,  where ${\cal J}_o$  
is the in-plane SC stiffness.  

We can make a crude estimate of the magnitude of the residual interplane couplings by noting that the same interplane matrix elements (although not necessarily the same energy denominators) determine the interplane exchange couplings between spins and the interplane Josephson couplings.  Defining $J_m$ to be the exchange couplings between % nearest-neighbor 
 spins $m$ planes apart, this estimate suggests that ${\cal J}_{n,m}/{\cal J}_0 \sim [J_m/J_0]^n$.  In undoped {\LCO}, it has been determined
\cite{bibbob} that $J_1/J_0\approx 10^{-5}$, which is already remarkably small.

Although in-plane translation invariance forbids direct Josephson coupling between adjacent planes, there is an allowed biquadratic  inter-plane coupling 
involving {\bf M} and $\Delta$, the SDW and the SC order parameters,
\begin{equation}
\delta H_{\rm inter}={\cal J}_{1,s}\sum_j \int d\vec{r}\; \left[ \Delta_j^* \Delta_{j+1} {\bf M}_j \cdot {\bf M}_{j+1}+{\rm h.c.} \right]
\end{equation}
Even though ${\bf M}\neq 0 $ for  $T < T_{spin}$, this term vanishes because, not only the direction of the stripes, but also the axis of quantization of the spins (due to 
 spin-orbit coupling) rotates \cite{hucker05}  by 
$90^\circ$ from plane to plane, {\it i.e.} ${\bf M}_j \cdot {\bf M}_{j+1}=0$.
However,
a magnetic field, $H \sim 6 T$, 
induces a 
 $1$st order spin-flop transition to a fully collinear spin state \cite{hucker05} in which ${\bf M}_j \cdot {\bf M}_{j+1}\neq 0$.

Thus, for perfect stripe order, the anti-phase SC order would 
depress, by many orders of magnitude, 
of 
the interplane Josephson couplings, which 
explains the existence of a broad range of $T$ in which 2D physics is apparent.  Accordingly,
there still would be a transition to a 3D superconductor at a temperature strictly greater than $T_{KT}$, when $\xi(T)\sim
\xi_{1,4}$ or $\xi_{2,1}$, whichever is smaller.  The only evidence for the growth of $\xi$ comes indirectly from the measurement of $\rho_{ab}$;  by the time $\rho_{ab}$ is ``unmeasurably small,'' it has dropped by about 2 orders of magnitude from its value just below $T_{\rm spin}$, which implies (since $\rho_{ab} \sim \xi^{-2}$) that $\xi$ has grown by about 1 order of magnitude.  Thus, if some other physics cuts off the growth of in-plane SC correlations at long scales, 
we may be justified in neglecting the effects of $H_{\rm inter}$.

Defects in the pattern of charge stripe order have 
consequences for both magnetic and SC orders.  
A dislocation
introduces frustration into the in-plane ordering, 
resulting in the formation of a half-SC vortex bound to it.
For the single-plane problem, this means that the long-distance physics is that of an XY spin-glass.  Since
there is no finite $T$ glass transition in 2D, 
the growth of $\xi$ will
be arrested 
at a large scale determined by the density of dislocations.  The same is true of the in-plane AFM correlations.  
Both $\xi$ and $\xi_{\rm spin}$ should be bounded above by 
the charge stripe correlation length, $\xi_{\rm ch}$.
From X-ray scattering studies it is estimated that 
$\xi_{\rm ch} \approx 70a$ \cite{kim07}.  This 
justifies the neglect of $H_{\rm inter}$.  Conversely, any defect in the charge-stripe order spoils the symmetry responsible for the exact cancellation of the Josephson coupling between neighboring planes. 
Finite $T$ ordering of an XY spin-glass is possible in 3D. We tentatively identify the temperature at which $\rho_c\to 0$ as a 3D glass transition.   
A SC glass
would result in the existence of equilibrium currents (spontaneous time-reversal breaking)
and
in glassy long-time relaxations of 
the magnetization or $\rho_c$.

For $x \neq 1/8$, there is a tendency to develop discommensurations in the stripe order, which, in turn, produce regions of enhanced (or  depressed) SC order with relative sign depending on the number of intervening stripe periods. 
So long as the stripes are dilute, the energy depends 
weakly on their precise spacing.  Thus, to gain interlayer condensation energy, the system can self-organize so that there are always an even number of intervening stripes, thus producing an interplane Josephson coupling ${\cal J}_{1,1} \sim |x - 1/8|^2$.  This, in turn, will lead to
%SAKa rapid suppression of the 2D KT regime and 
a dramatic increase of the 3D SC $T_c$.
An enhancement of interplane coherence in any range of $T$ triggered by the magnetic field induced spin-flop transition would be a dramatic confirmation of the physics discussed here.\\
{\em Note added:} It was pointed out to us that the state discussed here was considered  by A. Himeda {\it et al}.\cite{ogata02} They found that this is a good variational state for a $t-t^\prime-J$ model at $x\sim 1/8$ for a narrow range of parameters.

We thank P. Abbamonte, S. Chakravarty, R. Jamei, A. Kapitulnik, and D. J. Scalapino for discussions.
This work was supported in part by the National Science Foundation, under grants DMR 0442537 (EF), DMR 0531196 (SAK), DMR 0342832 (SCZ), and  by the Office
of Science, U.S. Department of Energy under Contracts DE-FG02-91ER45439 (EF), 
DE-FG02-06ER46287 (SAK)
DE-AC02-98CH10886 (JT) and DE-AC03-76SF00515 (SCZ), by the Stanford Institute for Theoretical Physics (EAK), and by a Yale Postdoctoral Prize Fellowship (VO).


\begin{thebibliography}{99}
\bibitem{bandm} J. G. Bednorz and K. A. Mueller, Z. Phys. B: Condens. Matter {\bf 64}, 189 (1986).
\bibitem{modenbaugh}  A. R. Modenbaugh {\it et al.}, \prb\ {\bf 38}, 4596 (1988).
\bibitem{tranquada}  J. M. Tranquada {\it et al.}, Nature, {\bf 375}, 561 
(1995).
\bibitem{abbamonte}  P. Abbamonte {\it et al.}, Nature Phys. {\bf 1}, 155 (2005).
\bibitem{newtranq}  Q. Li, M. H{\"u}cker, G. D. Gu, A. M. Tsvelik, and J.
M. Tranquada, cond-mat/0703357.
\bibitem{valla}  T. Valla {\it et al}, Science {\bf 314}, 1914 (2006).
\bibitem{zimmerman} M. v. Zimmermann {\it et al}, Europhys. Lett. {\bf
41}, 629 (1998).
\bibitem{fujita}  M. Fujita {\it et al.}, \prb\ {\bf 70}, 104517 (2004).  In 1/8 doped {\LNSCO}, J. M. Tranquada {\it et al.}, 
\prb\ {\bf 59}, 14712 (1999) found $\xi_{\rm spin} \approx 50a$ (with no substantial interplane correlations). In stage IV O-doped {\LCO} (which does not exhibit the LTT structure),  
Y. S. Lee {\it et al.}, \prb\ {\bf 60}, 3643 (1999), found  
$\xi_{\rm spin}
>100a$
with c-axis correlations
of 
2--3 planes.
\bibitem{ekz}  V. J. Emery,  S. A. Kivelson and O. Zachar, \prb\ {\bf 56}, 6120 (1997).
\bibitem{afk} E. Arrigoni, E. Fradkin, and S. A. Kivelson, \prb\ {\bf 69}, 214519 (2004).
\bibitem{getal} M. Granath {\it et al.}, \prl\ {\bf 87}, 167011 (2001).
\bibitem{8leg}  S. R. White and D. J. Scalapino, \prl\ {\bf 80}, 1272 (1998).
\bibitem{white2leg}  S. R. White, I. K. Affleck and D. J. Scalapino, \prb\ {\bf 65}, 165122 (2002). 
\bibitem{dwave} A ``d-wave like'' gap changes sign as well as magnitude under rotation by $\pi/2$.
\bibitem{davis}  Y. Kohsaka {\it et al.}, Science {\bf 315}, 1380 (2007).
\bibitem{3leg} S. R. White and D. J. Scalapino, \prb\ {\bf 57}, 3031 (1998).
\bibitem{zachar}  O. Zachar, \prb\ {\bf 65}, 174411 (2002).
\bibitem{pryadko}  L. Pryadko {\it et al.}, \prb\ {\bf 60}, 7541 (1999).
\bibitem{liu01} W. V. Liu and E. Fradkin \prl\ {\bf 86}, 1865 (2001).
\bibitem{spivak}  B. I. Spivak and S. A. Kivelson,\prb\ {\bf 43}, 3740 (1991).
\bibitem{zhang} S-C. Zhang, J. Phys. Chem. Solids {\bf 59}, 1774 (1998).
\bibitem{bibbob}
B. Keimer {\it et al.}, Phys.\ Rev.\ B {\bf 46}, 14034 (1992).
\bibitem{hucker05} M. H\"ucker, G. Gu and J. M. Tranquada, cond-mat/0503417.
\bibitem{kim07} J. Kim, A. Kagedan, G. D. Gu, C. S. Nelson, T. Gog, D.
Casa, and Y.-J. Kim, cond-mat/0703265.
\bibitem{ogata02} A. Himeda, T. Kato and M. Ogata, \prl\ {\bf 88}, 117001 (2002).
\end{thebibliography}
\end{document}